\documentclass[reprint, aps,prd, preprintnumbers,groupedaddress,nofootinbib]{revtex4-1}
\usepackage{graphicx}
\usepackage{latexsym}
\usepackage{amsfonts}
\usepackage{amssymb}
\usepackage{amsmath}
\usepackage{slashed}
\usepackage{feynmp}
\usepackage{hyperref}
\usepackage{url}
\usepackage{color}
\usepackage{cancel}
\usepackage[normalem]{ulem}
\usepackage{multirow,array}
\usepackage{dcolumn}
\usepackage{bm}
\usepackage{tabu}

\newcommand\one{\leavevmode\hbox{\small1\normalsize\kern-.33em1}}

\def\slashchar#1{\setbox0=\hbox{$#1$}           
   \dimen0=\wd0                                 
   \setbox1=\hbox{/} \dimen1=\wd1               
   \ifdim\dimen0>\dimen1                        
      \rlap{\hbox to \dimen0{\hfil/\hfil}}      
      #1                                        
   \else                                        
      \rlap{\hbox to \dimen1{\hfil$#1$\hfil}}   
      /                                         
   \fi}

\newcommand{\be}{\begin{eqnarray*}}
\newcommand{\ee}{\end{eqnarray*}}

\newcommand{\bee}{\begin{eqnarray}}
\newcommand{\eee}{\end{eqnarray}}
\newcommand{\beeq}{\begin{equation}}
\newcommand{\eeeq}{\end{equation}}

\usepackage{tikz}
\usepackage[compat=1.1.0]{tikz-feynman}

\definecolor{orangeRMA}{RGB}{255,127,0}

\begin{document}

\title{Boosting New Physics Searches in $t\bar{t}Z$ and $tZj$ Production with Angular Moments}

\author{Roshan Mammen Abraham}
\author{Dorival Gon\c{c}alves} 
\affiliation{Department of Physics, Oklahoma State University, Stillwater, OK, 74078, USA}


\begin{abstract}

The angular moments of the $Z$ boson can be used  as analyzers for the underlying production dynamics for the $t\bar t Z$ and $t Zj$  processes. In this manuscript, we derive these angular moments at leading and next-to-leading order in QCD at the LHC.  We show that these observables work as efficient probes to beyond the Standard Model effects, considering the Standard Model Effective Field theory framework. Remarkably, we observe that these  probes unveil blind directions to CP-odd operators, providing sizable new physics sensitivity at the 14~TeV LHC with 3~ab$^{-1}$ of data.
\end{abstract}

\maketitle

\section{Introduction}

Precision studies for top quark physics are a cornerstone for the LHC program. The large top quark mass indicates that it  may have a special role in electroweak symmetry breaking (EWSB)~\cite{Carena:1993bs,PhysRevD.75.015002,PhysRevD.80.076002,Panico:2011pw,Matsedonskyi:2012ym,Pomarol:2012qf,PhysRevD.84.015003,Bellazzini:2014yua,Panico:2015jxa}. Thus, top quark precision measurements can display the first glimpse into new physics connected with EWSB.  While the basic top quark properties (\emph{e.g.}, mass, pair production cross-section, and $W$-helicity fractions) are well known and consistent with the Standard Model (SM)~\cite{10.1093/ptep/ptaa104}, its interaction with the $Z$ boson is still weakly constrained. 

The most promising  \emph{direct} probes for the top quark-$Z$ boson interaction are via production at the LHC of a top pair and a $Z$ boson $pp\to t\bar{t}Z$ and  single top production in association with a $Z$ boson and a jet $pp\to tZj$~\cite{Baur:2004uw,Kidonakis:2022ljg}. The large production threshold of $2m_t+m_Z$ for $t\bar t Z$  and the small electroweak production rate for $tZj$  require the sizable collision energy and luminosity provided by the LHC, making these  probes unattainable at previous colliders. The most recent experimental measurements for the top quark-$Z$ boson interaction are reported by  ATLAS with $139.1~\text{fb}^{-1}$~\cite{ATLAS:2021fzm} and CMS with $77.5~\text{fb}^{-1}$~\cite{CMS:2019too},  displaying good agreement with the theoretical calculations within the SM.  Experimental projections indicate that the top quark electroweak interaction will be probed to great precision when going from the Run 2 dataset of 139~fb$^{-1}$ to the projected high luminosity LHC (HL-LHC) with 3~ab$^{-1}$~\cite{CMS-PAS-FTR-18-036}. These analyses can ultimately shed light on  well motivated connections of the top quark to new physics.  
 
In the present study, we show the possibility to boost the new physics sensitivity in the $t\bar tZ$ and $tZj/\bar{t}Zj$ processes  at the LHC using the angular moments for the $Z$ boson~\cite{PhysRevLett.52.1076,Mirkes:1994eb,Goncalves:2018fvn,Goncalves:2018ptp,Banerjee:2019pks,Banerjee:2020vtm,Banerjee:2019twi}.  This proposal scrutinizes the hadronic structure of the processes under inspection through the full $Z$ boson polarization information, using the leptons as spin analyzers for the underlying production dynamics. While this phenomenological probe is disregarded in the current experimental analyses, we show that the proposed method can be a key ingredient to access new physics contributions at higher precision.

We parametrize new physics effects in terms of the SM Effective Field theory (EFT) framework~\cite{Buchmuller:1985jz,Grzadkowski:2010es,Brivio:2017vri}. The EFT provides a well-defined approach to explore indirect effects from new theories as deformations from the SM structures. These new physics effects would generally manifest as subtle deviations in the standard physics observables. 

The paper is organized as follows. In Section~\ref{sec:frame}, we present the SM angular moments for the $t\bar{t}Z$ and $tZj/\bar{t}Zj$ processes and quantify the higher order QCD effects. In Section~\ref{sec:eft}, we present the relevant operators in the EFT framework up to dimension-six and calculate their new physics contributions to the  observables under scrutiny. In Section~\ref{sec:ana}, we show our detector level analysis and discuss the HL-LHC sensitivity to the corresponding Wilson coefficients. We draw our conclusion in Section~\ref{sec:conclusion}.

\section{Theoretical Framework}
\label{sec:frame}

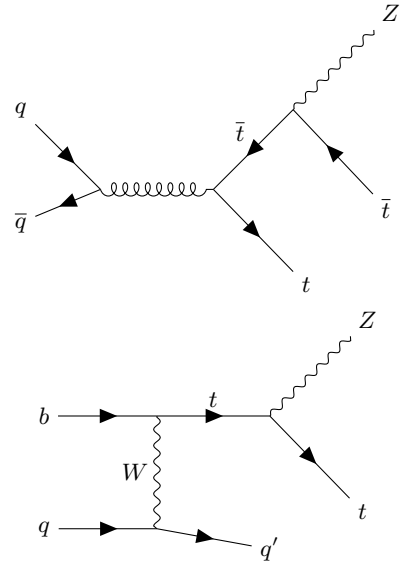
\begin{figure}[b]
    \centering
    
    \centering

    \begin{tikzpicture}
    \begin{feynman}
    
    \vertex (i1) {\(q\)};
    \vertex [below = of i1] (i2) {\(\overline{q}\)};
    \vertex [below right = of i1] (a);
    \vertex [right = of a] (b);
    \vertex [above right = of b] (c);
    \vertex [below right = of b] (f3) {\(t\)};
    \vertex [above right = of c] (f1) {\(Z\)};
    \vertex [below right = of c] (f2) {\(\overline{t}\)};
    
    \diagram*{
    
    (i1) -- [fermion] (a) -- [fermion] (i2),
    (a) -- [gluon] (b),
    (f2) -- [fermion] (c) -- [fermion, edge label'=\(\overline{t}\)] (b) -- [fermion] (f3),
    (c) -- [boson] (f1),
    
    };
    
    \end{feynman}
    \end{tikzpicture}

    \centering

    \begin{tikzpicture}
    \begin{feynman}
    
    \vertex (i1) {\(b\)};
    \vertex [below = of i1] (i2) {\(q\)};
    \vertex [right = of i1] (a);
    \vertex [below = of a] (b);
    \vertex [right = of a] (c);
    \vertex [below = of c] (f3) {\(q'\)};
    \vertex [above right = of c] (f1) {\(Z\)};
    \vertex [below right = of c] (f2) {\(t\)};
    
    \diagram* {
    
    (i1) -- [fermion] (a) -- [fermion, edge label=\(t\)] (c) -- [fermion] (f2),
    (i2) -- [fermion] (b) -- [fermion] (f3),
    (a) -- [ boson, edge label'=\(W\)] (b),
    (c) -- [photon] (f1),
    
    };
 
    \end{feynman}
    \end{tikzpicture}
    \caption{Representative set of Feynman diagrams for the $pp\to t\bar t Z$ (top) and $pp\to tZj$  (bottom) processes.}
    \label{fig:FeynDiag}
\end{figure}

\begin{figure*}[t]
\centering
\includegraphics[width=0.45\textwidth]{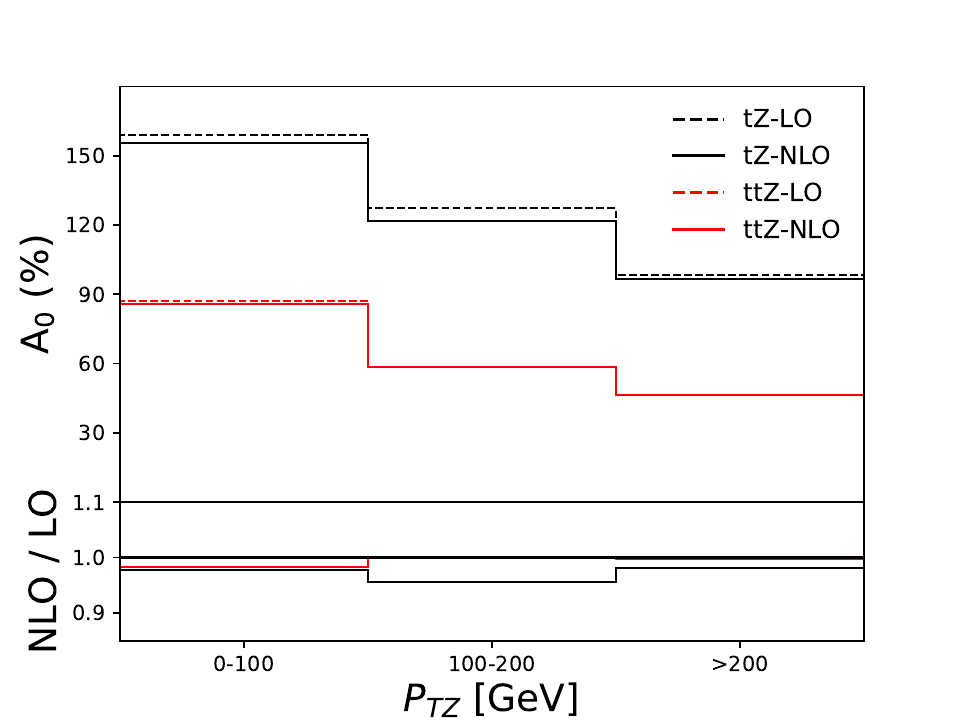}%
\hspace{1cm}
\includegraphics[width=0.45\textwidth]{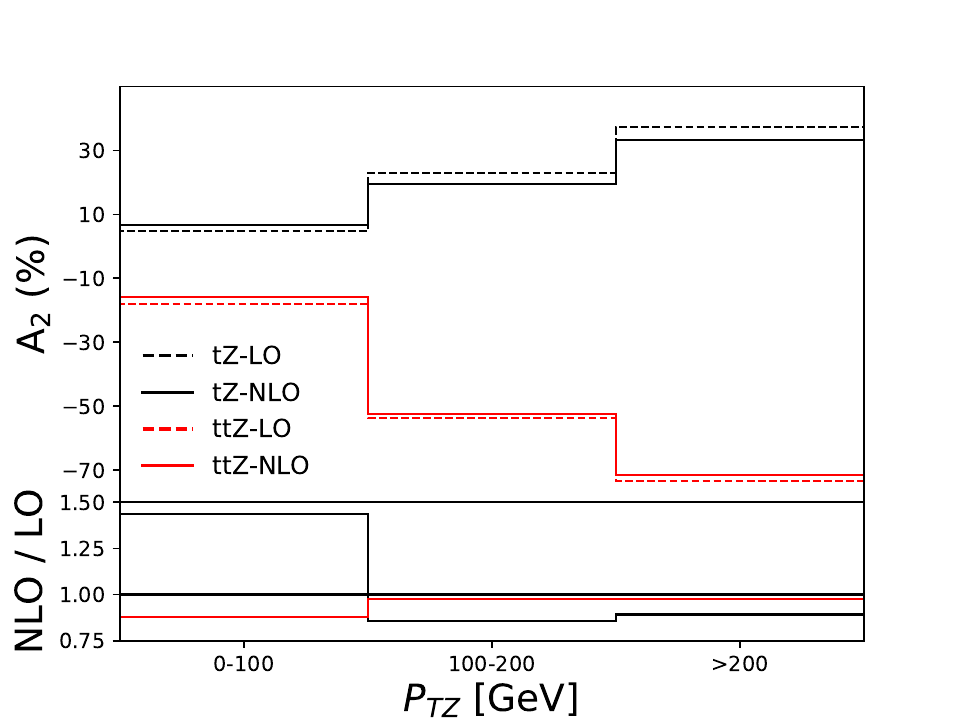}%
\caption{Angular coefficients $A_0$ (left panel) and $A_2$ (right panel) for top quark pair plus dilepton $pp\rightarrow t\bar{t}\ell^+\ell^-$ (red) and single top quark plus dilepton $pp\rightarrow t(\bar{t})\ell^+\ell^-j$ (black). The results are presented at LO (dashed line) and NLO (solid line). The processes are calculated at the parton level with $|\eta_\ell|<4$, $p_{T\ell}>5$~GeV, and  $|m_{\ell\ell}-m_Z|<10$~GeV. The renormalization and factorization scales are set to  $\mu_R=\mu_F=1/2\sum_{i=1}^{n}\sqrt{m_i^2+p_{T,i}^2}$.
\label{fig:a02SM}}
\end{figure*}

In the present manuscript, we show that the  angular distribution in the $Z\rightarrow \ell^+\ell^-$ decay opens a gateway for precision studies in the $pp\rightarrow t \bar{t}Z$ and $tZj/\bar{t}Zj$  processes.  In general, the differential cross-section for these processes can be written as~\cite{PhysRevLett.52.1076,Mirkes:1994eb}
\begin{align}
& \frac{1}{\sigma}\frac{d\sigma}{d\cos{\theta}d\phi} =\nonumber \\
& \frac{3}{16\pi} [ 1+\cos^2{\theta} 
+ A_{0} \frac{1}{2} (1-3\cos^2{\theta} )
+ A_{1} \sin{2\theta} \cos{\phi}    \nonumber  \\
& + A_{2}\frac{1}{2} \sin^2{\theta} \cos{2\phi} 
+A_{3} \sin{\theta} \cos{\phi} 
+A_{4} \cos{\theta} \nonumber  \\
&+A_{5} \sin^2{\theta} \sin{2\phi}
+ A_{6} \sin{2\theta} \sin{\phi} 
+ A_{7} \sin{\theta} \sin{\phi}]  \,,
\label{differential}
\end{align}
where $\theta$ and $\phi$ are the polar and azimuthal angles  of the  $\ell^-$ lepton in the $Z$ boson rest frame.  The eight coefficients $A_i$, $i=[0,7]$, correspond to the number of degrees of freedom for the polarization density matrix for a spin-1 particle. The angular coefficients $A_i$ are frame dependent. We adopt the Collins-Soper frame in our study~\cite{PhysRevD.16.2219}. This is a typical frame choice in angular coefficient analyzes~\cite{ATLAS:2016rnf,ATLAS:2017rue,CMS:2015cyj}.

Our studies will focus on the top quark and $Z$ boson interaction via top quark pair production in association with a $Z$ boson  $pp\rightarrow t\bar{t}Z$  and single top quark production in association with a $Z$ boson and a jet  $pp\rightarrow tZj/\bar{t}Zj$.  See Fig.~\ref{fig:FeynDiag} for a representative set of Feynman diagrams. We consider the semi-leptonic top pair decays and $Z\rightarrow \ell^+\ell^-$. The Monte Carlo analysis sums over all possible combinations of charged leptons $\ell^\pm =e^\pm, \mu^\pm$. Before analyzing the angular coefficients in the quest for new physics, we study in this section the stability of these terms to higher order effects.

Event generation for $pp\rightarrow t\bar{t}\ell^+ \ell^-$ and $pp\rightarrow t(\bar{t})\ell^+ \ell^-j$ processes is performed at leading order (LO) and next to leading order (NLO) QCD with MadGraph5\_aMC@NLO~\cite{Alwall:2014hca}. We consider the LHC at $\sqrt{s}=14$~TeV. Both the $Z$ and $\gamma^*$ intermediate states, associated to the dilepton final state, are accounted for. To isolate the higher order effects in our simulation, we perform a parton level study in this section, requiring only basic selections to the two charged leptons from the $Z/\gamma^*$ decays, keeping the top quark pair stable. Leptons are defined with $|\eta_\ell|<4$ and $p_{T\ell}>5$~GeV. We demand a charged lepton pair, with same flavor and opposite sign, reconstructing the $Z$ boson mass $|m_{\ell\ell}-m_Z|<10$~GeV. The renormalization and factorization scales are dynamically defined as $\mu_R=\mu_F=1/2\sum_{i=1}^{n}\sqrt{m_i^2+p_{T,i}^2}$. We adopt the parton distribution function NNPDF23 at NLO with $\alpha_s(m_Z)=0.119$~\cite{Ball:2013hta}.

To extract the angular coefficients $A_i$ from our Monte Carlo simulation, we observe that Eq.~(\ref{differential}) is a spherical harmonic decomposition for the differential cross-section with real spherical harmonics $Y_{lm}(\theta,\phi)$ of order $l\le 2$. Hence, we can access the angular coefficients, exploring the orthogonality relations for the spherical harmonics. The angular coefficients are projected out with
\begin{align}
&A_0=4-\left <10\cos^2\theta \right >,         &A_1&=\left <5\sin 2\theta\cos\phi \right > ,  \nonumber \\
&A_2=\left <10\sin^2\theta\cos 2\phi\right > ,  &A_3&=\left<4\sin\theta\cos\phi  \right> , \nonumber \\
&A_4=\left <4\cos\theta \right > ,              &A_5&=\left<5\sin^2\theta\sin 2\phi  \right> , \nonumber \\
&A_6=\left <5\sin 2\theta\sin\phi \right > ,    &A_7&=\left<4\sin\theta\sin\phi \right> , 
\label{eq:coef}
\end{align}
where the weighted normalization is defined as
\begin{align}
\left <f(\theta,\phi)\right >\equiv\int_{-1}^{1} d\cos\theta \int_0^{2\pi}d\phi  f(\theta,\phi) \frac{1}{\sigma}\frac{d\sigma}{d\cos\theta d\phi}\,.
\label{eq:weight}
\end{align}
In this definition, $\sigma$ can represent any differential cross-section that is independent of the lepton kinematics.

\begin{table*}[t!] 
\begin{center}
    \begin{tabular}{c c c c c c c c c}
    \hline\hline
    & $A_0$ & $A_1$ & $A_2$ & $A_3$ & $A_4$ & $A_5$ & $A_6$ & $A_7$ \\
    \hline
    $t\bar{t}Z_\text{LO}$ & 0.693(9) & 0.004(9) & -0.41(1) & 0.00(0.01) & 0.00(1) & 0.010(9) & 0.00(1) & 0.00(1) \\
    $t\bar{t}Z_\text{NLO}$ &  0.68(1) &  -0.003(7) & -0.39(1) & 0.004(4) & 0.001(7) & 0.001(2) & 0.000(6) & 0.000(2) \\
    $t(\bar{t})Z_\text{LO}$ & 1.46(2) & 0.001(8) & 0.117(9) & 0.04(1) & 0.000(8) & -0.003(9) & 0.001(8) & 0.00(1) \\
     $t(\bar{t})Z_\text{NLO}$ & 1.41(1) & -0.008(5) & 0.12(1) & 0.035(6) & -0.005(6) & -0.002(6) & -0.006(7) & -0.001(9) \\
    \hline \hline
    \end{tabular}
    \caption{Angular coefficients $A_i$ for top pair plus dilepton $pp\rightarrow t\bar{t}\ell^+\ell^-$  and single top plus dilepton $pp\rightarrow t(\bar{t})\ell^+\ell^-j$  processes at LO and NLO QCD. The processes are calculated at the parton level with $|\eta_\ell|<4$, $p_{T\ell}>5$~GeV, and $|m_{\ell\ell}-m_Z|<10$~GeV. The Monte Carlo statistical uncertainty, represented as one standard deviation, is enclosed within parentheses for the last digit.
    \label{tab:ang-mom}}
\end{center}
\end{table*}

The properties of these angular coefficients can provide valuable insights into physics. Among them, the three coefficients $A_{5-7}$ display direct proportionality to the relative complex phases of the scattering amplitudes~\cite{Goncalves:2018fvn,Aguilar-Saavedra:2017zkn}. Consequently, when these coefficients are linked with reduced strong phase contributions arising from loop processes, they become particularly sensitive to genuine CP-violation effects.
    Furthermore, the the coefficients $A_{3,4,7}$ are proportional to the polarization analyzer $\eta_\ell = \left( (g_L^\ell)^2-(g_R^\ell)^2 \right)/\left((g_L^{\ell})^2+(g_R^{\ell})^2\right) \approx 0.14$~\cite{Aguilar-Saavedra:2017zkn}. The inherently small value of $\eta_\ell$ naturally accounts for the depleted magnitude in the coefficients $A_{3,4,7}$. Nevertheless, we take a comprehensive approach and numerically calculate all the $A_i$ coefficients in our study.

In Table~\ref{tab:ang-mom}, we present the angular coefficients $A_i$ at LO and NLO QCD. We observe that the angular distributions for the leptons are controlled by two leading terms, namely $A_0$ and $A_2$. The higher order corrections display relevant dependencies with the $Z$ boson transverse momentum, see Fig.~\ref{fig:a02SM}. The other angular coefficients result in sub-leading effects.

 
\section{Effective Field Theory}
 \label{sec:eft}

The current LHC constraints point to a mass gap between the SM degrees of freedom and the new physics states. In this context, the new physics modes can be integrated out and be well parametrized by high dimension operators within the SM Effective Field Theory  framework~\cite{Appelquist:1974tg,Buchmuller:1985jz,Grzadkowski:2010es,Brivio:2017vri}. In the present section, we study the effects of higher dimensional operators that  influence the interaction between the top quark and neutral gauge bosons and are relatively unconstrained~\cite{ttZ_coupling,Rontsch:2015una,Buckley:2015lku,Schulze:2016qas,BessidskaiaBylund:2016jvp,CMS:2017ugv,Englert:2017dev,Degrande:2018fog,CMS:2019too,Maltoni:2019aot,Brivio:2019ius,Durieux:2019rbz,ATLAS:2021fzm,Ravina:2021kpr,Rahaman:2022dwp,Barman:2022vjd}. Following the Warsaw basis~\cite{Grzadkowski:2010es}, we focus on the operators
\begin{align}
&\mathcal{O}_{tB} = \left(\overline{Q}\sigma^{\mu \nu}t\right)\widetilde{\phi}B_{\mu \nu}  \,,  \nonumber \\
&\mathcal{O}_{tW} = \left(\overline{Q}\sigma^{\mu \nu}\tau^{I}t\right)\widetilde{\phi}W^{I}_{\mu \nu}   \,,  \nonumber \\
&\mathcal{O}_{\phi t}=\left(\phi^{\dagger}i\overleftrightarrow{D_{\mu}}\phi\right)\left(\overline{t}\gamma^{\mu}t\right)  \,, \nonumber \\
&\mathcal{O}^{\left(1\right)}_{\phi Q}= \left(\phi^{\dagger}i\overleftrightarrow{D_{\mu}}\phi\right)\left(\overline{Q}\gamma^{\mu}Q\right)\,, \nonumber \\
&\mathcal{O}^{\left(3\right)}_{\phi Q}=\left(\phi^{\dagger}i\overleftrightarrow{D_{\mu}}\tau^{I}\phi\right)\left(\overline{Q}\gamma^{\mu}\tau^{I}Q\right)\,,
\label{eq:ope}
\end{align}
where $Q$ denotes the left-handed top-bottom doublet and $t$ the right-handed top singlet. $\tau^I$ are the Pauli matrices, and the Higgs doublet is represented by $\phi$ and $\tilde{\phi}\equiv i\tau^2 \phi$.

The BSM contributions to the top quark and $Z$ boson interaction can be parametrized by the Wilson coefficients $(c_{\phi t},c_{tZ},c_{tZ}^I,c_{\phi Q})$. The last three coefficients are defined from the following linear combinations~\cite{Aguilar-Saavedra:2018ksv,CMS:2019too}
\begin{align}
    c_{tZ}&\equiv\mbox{Re}\left( -\sin \theta_W c_{tB}+\cos\theta_W c_{tW} \right)\,,\\
       c_{tZ}^I&\equiv\mbox{Im}\left( -\sin \theta_W c_{tB}+\cos\theta_W c_{tW} \right)\,,\\
    c_{\phi Q}&\equiv c_{\phi Q}^1-c_{\phi Q}^3\,,
    \label{eq:cphiQ}
\end{align}
where $\theta_W$ is the Weinberg angle.\footnote{Several alternative theories that go beyond the Standard Model and focus on explaining the process of electroweak symmetry breaking propose significant couplings involving dipole moments. These theories also suggest modifications to the vector and axial couplings of $t\bar{t}Z$ interactions compared to their values in the SM~\cite{Hollik:1998vz,Agashe:2006wa,Kagan:2009bn,PhysRevD.82.055001,PhysRevD.84.015003}.}

Although we follow the EFT framework, it is illuminating to observe how these operators translate to the anomalous coupling approach~\cite{ttZ_coupling}. In this context, the possible effects from physics beyond the SM are modeled by the extended Lagrangian for the $t\bar{t}Z$ interaction
\begin{align}
\mathcal{L}_{t\bar{t}Z}=&
e\bar{u}(p_{t})[\gamma_{\mu}\left(C_{1,V}+\gamma_{5}C_{1,A}\right)\notag\\
&+\frac{i\sigma^{\mu \nu}q_{\nu}}{M_{Z}}\left(C_{2,V}+i\gamma_{5}C_{2,A}\right)]v(p_{\bar t})Z_{\mu}
\,,
\label{eq:ano}
\end{align}
where $e$ is the electromagnetic coupling constant, $q_\nu=(p_t - p_{\bar{t}})_\nu$, and $\sigma_{\mu\nu}=\frac{i}{2}[\gamma_\mu,\gamma_\nu]$. In the Standard Model, the vector and axial couplings are respectively $C_{1,V}^\text{SM}\approx 0.24$ and $C_{1,A}^\text{SM}\approx -0.60$. In addition, the weak magnetic $C_{2,V}$ and electric dipole $C_{2,A}$ interactions  are zero at tree level. Higher order corrections in the SM generate subleading contributions to these terms with $C_{2,V}\approx 10^{-4}$~\cite{Bernabeu:1995gs} and $C_{2,A}$ being further suppressed, appearing only at three-loops~\cite{Czarnecki:1996rx,Hollik:1998vz}. 

The EFT contributions in Eq.~\eqref{eq:ope}, which respect the SM symmetries, can be translated in terms of the anomalous couplings as~\cite{AguilarSaavedra:2008zc}
\begin{align}
C_{1,V}=&C_{1,V}^\text{SM}+\frac{v^2}{2\Lambda^2\sin\theta_W\cos\theta_W}\mbox{Re}\left[-c_{\phi t}+ \left( c_{\phi Q}^3-c_{\phi Q}^1 \right)   \right], \nonumber \\
C_{1,A}=&C_{1,A}^\text{SM}+\frac{v^2}{2\Lambda^2\sin\theta_W\cos\theta_W}\mbox{Re}\left[-c_{\phi t} -\left( c_{\phi Q}^3-c_{\phi Q}^1\right)   \right], \nonumber \\
C_{2,V}=&\frac{\sqrt{2}v^2}{2\Lambda^2\sin\theta_W\cos\theta_W}\mbox{Re}\left[-\sin{\theta_w}c_{tB} + \cos{\theta_w}c_{tW}\right], \nonumber \\
C_{2,A}=&\frac{\sqrt{2}v^2}{2\Lambda^2\sin\theta_W\cos\theta_W}\mbox{Im}\left[-\sin{\theta_w}c_{tB} + \cos{\theta_w}c_{tW}\right].
\end{align}
In this form, it can be seen that the Wilson coefficient $c_{tZ}$ generates the weak magnetic dipole moment and its imaginary counterpart $c_{tZ}^I$ sources the electric dipole moment. At the same time, the coefficients $c_{\phi t}$ and $c_{\phi Q}$ induce anomalous neutral current interactions. Remarkably, the Wilson coefficients $c_{\phi Q}^3$ and $c_{\phi Q}^1$ only appear with an opposite sign, hence the associated production of top quark(s) and $Z$ boson  ($t\bar t Z$ and $tZj/\bar{t}Zj$) can only constrain the coefficient $c_{\phi Q}$ defined in Eq.~\eqref{eq:cphiQ}.

\begin{figure}[!t]
\centering
\includegraphics[width=0.45\textwidth]{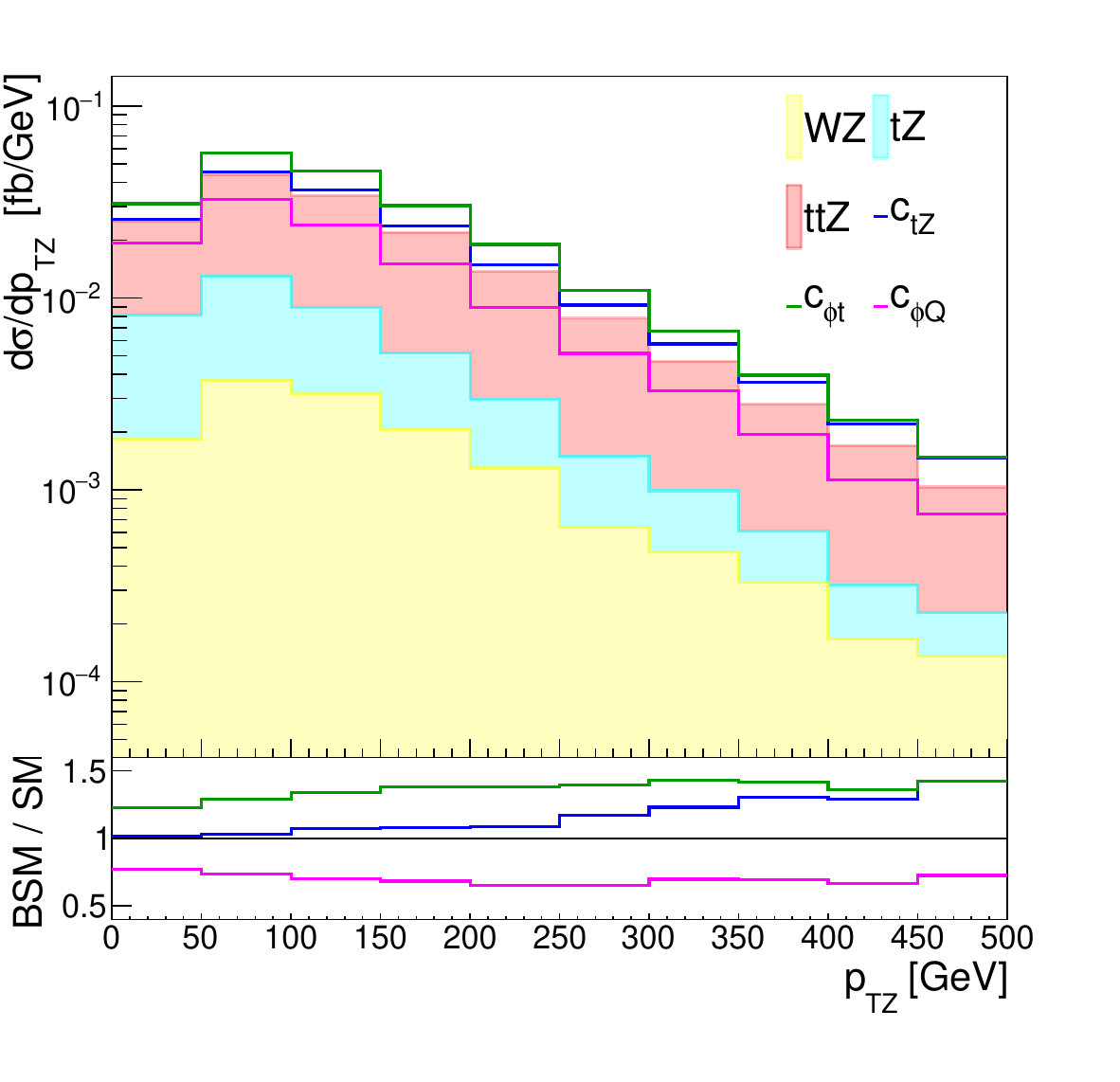}%
\caption{NLO differential cross-section as a function of $p_{TZ}$ for the SM and illustrative new physics scenarios. The Wilson coefficients are turned on one at a time to $c_{tZ}=1~\text{TeV}^{-2}$ and $c_{\phi t}=c_{\phi Q}=5~\text{TeV}^{-2}$. The new physics terms scale up to $\mathcal{O}(1/\Lambda^4)$ and the histograms are stacked. We show the ratio between the stacked BSM histograms and the SM in the bottom panel.}
\label{fig:sigmaPtZ}
\end{figure}

\begin{figure*}[t!]
\centering
\includegraphics[width=0.5\textwidth]{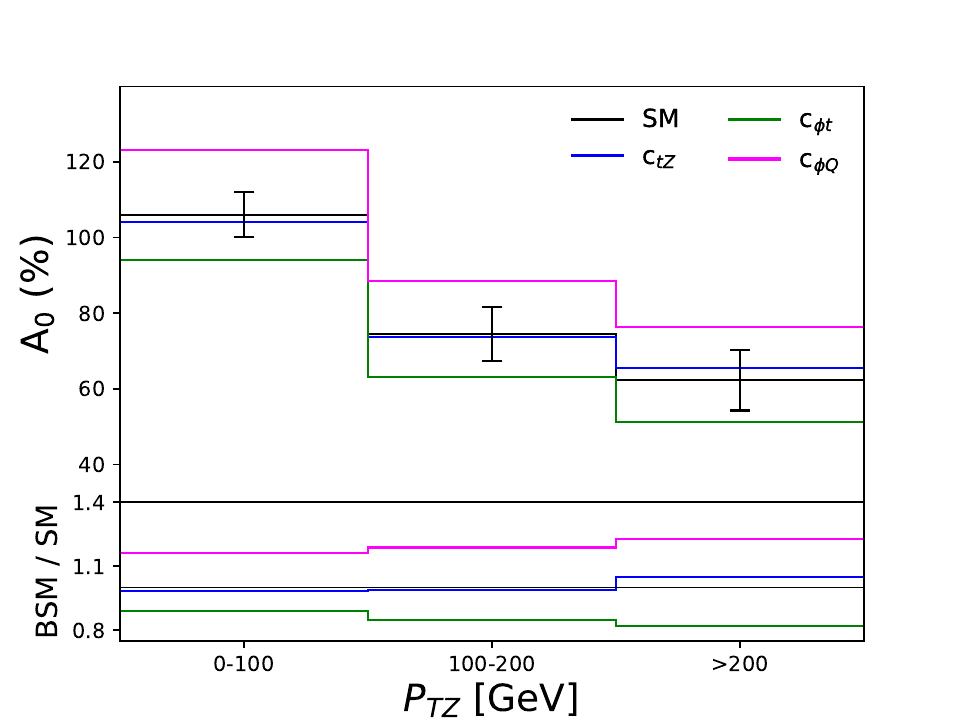}%
\includegraphics[width=0.5\textwidth]{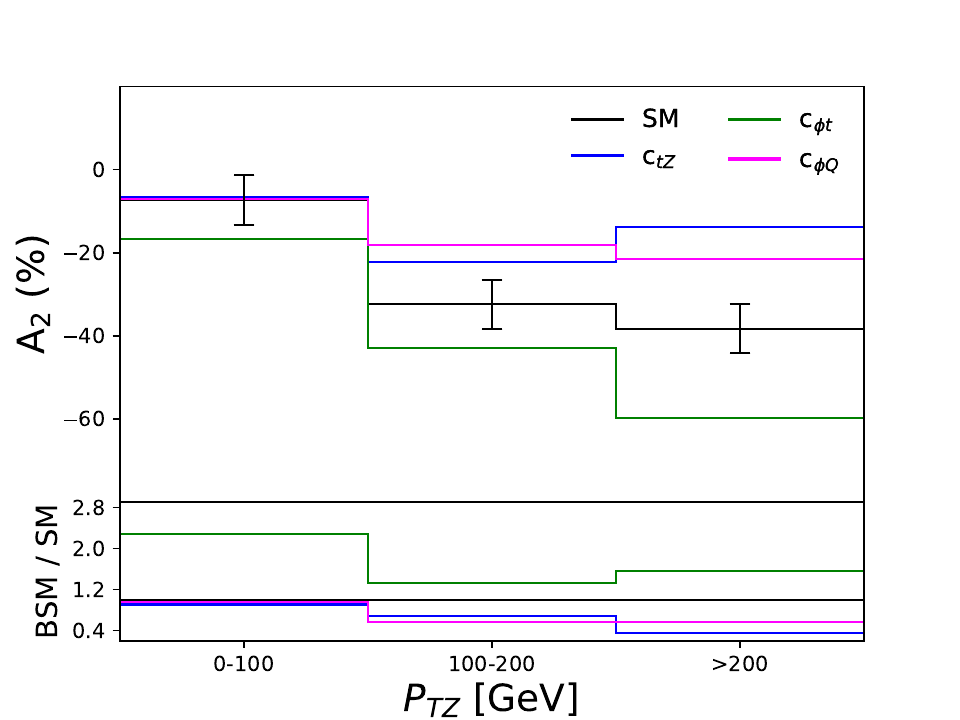}%
\caption{Angular coefficients $A_0$ (left panel) and $A_2$ (right panel) as a function of $p_{TZ}$ for the SM and new physics hypotheses for the combined samples $t\bar{t}Z$, $t(\bar{t})Z$, and $WZ$. The Wilson coefficients are turned on one at a time to $c_{tZ}=1~\text{TeV}^{-2}$ and $c_{\phi t}=c_{\phi Q}=5~\text{TeV}^{-2}$. 
The error bars represent the Monte Carlo statistical uncertainty of one standard deviation on the SM value, estimated through 100 pseudo-experiments as detailed in the text.
\label{fig:a02BSM}}
\end{figure*}

\begin{figure}[t!]
\centering
\includegraphics[width=0.5\textwidth]{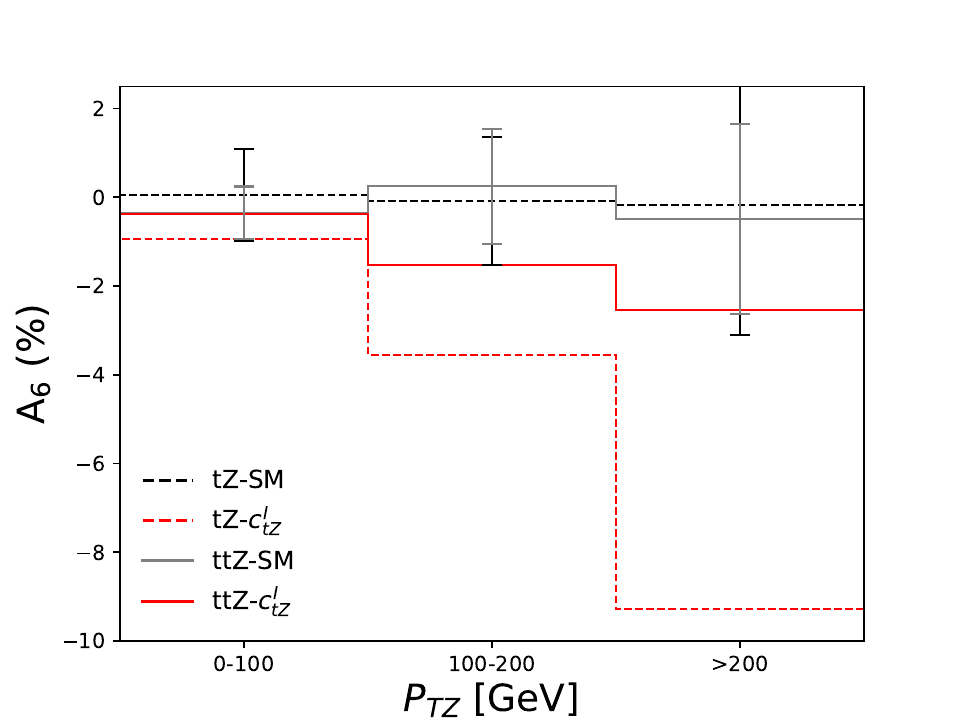}%
\caption{Angular coefficients $A_6$ as a function of the $Z$ boson transverse momentum $p_{TZ}$ for the SM (black, gray) and BSM CP-violating hypothesis $c_{tZ}^I$ (red). The results for the $t\bar t Z$ (solid) and $t(\bar{t})Z$ (dashed) processes are presented separately. The samples were generated at LO QCD and the Wilson coefficient is set to $c_{tZ}^I=1~\text{TeV}^{-2}$.
The error bars represent the Monte Carlo statistical uncertainty of one standard deviation on the SM values.
\label{fig:A6}}
\end{figure}

\section{Analysis}
\label{sec:ana}

In our analysis, we focus on the associated production of top quark(s) and a $Z$ boson ($ t\bar t Z$ and $tZj/\bar{t}Zj$), considering the final state with the $Z$ boson decaying leptonically and one top quark decaying semi-leptonically. To probe the HL-LHC sensitivity to new physics effects, we use MadGraph5\_aMC@NLO with the UFO model \texttt{SMEFTatNLO}~\cite{madgrpah,smeft}. This model file grants  EFT studies at NLO QCD for the CP-conserving operators $(c_{\phi t},c_{tZ},c_{\phi Q})$. The CP-violating contributions for $c_{tZ}^I$  are generated with the UFO model file \texttt{dim6top}~\cite{Aguilar-Saavedra:2018ksv}, that provides EFT samples at LO. Spin correlation effects for the top quark pair decays are obtained with  {\sc MadSpin} package~\cite{madpsin}. The leading background for this analysis arises from $WZ$ production, which is also simulated with MadGraph. Parton shower, hadronization, and underlying event effects are accounted for with {\sc Pythia8}~\cite{Sjostrand:2007gs}. Detector effects are simulated with {\sc Delphes3}~\cite{Ovyn:2009tx}, using  the  default  HL-LHC detector card~\cite{Cepeda:2019klc}. We consider the LHC at $\sqrt{s}=14$~TeV.

We start our detector level analysis, requiring three charged leptons. Leptons are defined with $|\eta_\ell|<4$ and $p_{T\ell}>5$~GeV. We demand a charged lepton pair, with same flavor and opposite sign, reconstructing the $Z$ boson mass $|m_{\ell\ell}-m_Z|<10$~GeV. For the hadronic part of the event, we require three or more jets where one is $b$-tagged. Jets are defined with the anti-$k_T$ jet algorithm with radius $R=0.4$, $|\eta_j|<4$, and $p_{Tj}>30$~GeV.

In Fig.~\ref{fig:sigmaPtZ}, we present the  NLO differential cross-section as a function of the reconstructed $Z$ boson transverse momentum for the  SM and CP-conserving EFT operators $(c_{\phi t},c_{tZ},c_{\phi Q})$. Remarkably, the $c_{tZ}$ contributions display augmented BSM effects at high energy scales. This can be understood by the extra momentum dependence arising from new physics. This is apparent, for instance, in the $C_{2,V}$ term of Eq.~\eqref{eq:ano}. In contrast, the other CP-conserving operators $(c_{\phi t},c_{\phi Q})$ result in almost constant corrections to the SM rate across all energy bins. 

\begin{table}[t!] 
\begin{center}
    \begin{tabular}{c c c c c | c c}
    \hline
    \hline
     & SM$_\text{NLO}$ & $c_{tZ}=1$ & $c_{\phi t}=5$ & $c_{\phi Q}=5$ & SM$_\text{LO}$ & $c_{tZ}^{I}=1$\\
    \hline
    $\sigma$ [fb] & 7.863 & 8.434 & 10.418 & 5.603 & 5.010& 5.349 \\
    $A_0$ & 0.803 & 0.788 & 0.521 & 0.976 & 0.886 & 0.892 \\
    $A_1$ & -0.003 & 0.001 & -0.002 & 0.001 & 0.001 & 0.000 \\
    $A_2$ & -0.265 & -0.198 & -0.459 & -0.160 & -0.226 & -0.179 \\
    $A_3$ & 0.009 & 0.014 & 0.004 & 0.015 & 0.015 & 0.013 \\
    $A_4$ & 0.000 & 0.000 & 0.000 & 0.000 & 0.000 & 0.001 \\
    $A_5$ & -0.001 & -0.001 & 0.002 & -0.002 & -0.001 & 0.000 \\
    $A_6$ & 0.000 & -0.003 & -0.003 & 0.001 & 0.000 & -0.013 \\
    $A_7$ & -0.001 & 0.000 & -0.002 & -0.004 & 0.000 & 0.000 \\
    \hline
    \hline
    \end{tabular}
\end{center}
\caption{Angular coefficient $A_i$ for the SM and new physics hypotheses. The results account for the combination of all leading channel contributions:  $pp\rightarrow t\bar{t}\ell^+\ell^-$, $pp\rightarrow t(\bar{t})\ell^+\ell^-$, and $WZ$.   The Monte Carlo events are generated at NLO QCD for the CP-conserving operators $(c_{\phi t},c_{tZ},c_{\phi Q})$ and LO for the CP-violating one $(c_{tZ}^I)$. The event generation includes parton shower, hadronization, and detector level effects. See the text for more details. The Wilson coefficients are  turned on one at a time with the following strengths: $c_{tZ}=c_{tZ}^{I}=1~\text{TeV}^{-2}$ and $c_{\phi t}=c_{\phi Q}=5~\text{TeV}^{-2}$. The new physics terms scale up to $\mathcal{O}(1/\Lambda^4).$
\label{tab:ang-mom-BSM}}
\end{table} 

The angular coefficients provide an extra phenomenological probe to these new physics effects. They work as  spin analyzers for the hadronic structure. 
In Table~\ref{tab:ang-mom-BSM}, we display the angular coefficients $A_{i}$  for the SM and new physics scenarios. To illustrate the distinctive BSM effects to the angular coefficients, we turn one  Wilson coefficient at a time with  strengths $c_{tZ}=c_{tZ}^I=1~\text{TeV}^{-2}$ and $c_{\phi t}=c_{\phi Q}=5~\text{TeV}^{-2}$. The two leading angular coefficients that control the angular distributions in the SM, $A_0$ and $A_2$, present large BSM effects for the considered deformations in the EFT parameter space. Furthermore, while the SM and CP-conserving operators display depleted angular coefficient $A_6$, being zero at tree level, the CP-violating operator $c_{tZ}^I$ presents a sizable contribution. The angular coefficient $A_6$ is sensitive to the imaginary part of the amplitude, arising from the CP-violating operator. In Figs.~\ref{fig:a02BSM} and~\ref{fig:A6}, we show that these angular coefficients result in relevant dependencies with the energy scale $p_{TZ}$. In particular,  we observe augmented BSM contributions in the boosted regime for the $c_{tZ}^I$ operator in Fig.~\ref{fig:A6}. The uplifted new physics effects at high scales appear for both the $t\bar t Z$ and $tZj/\bar{t}Zj$ processes, being more pronounced for the latter.

To evaluate the sensitivity of these new BSM probes, we perform a bin-by-bin $\chi^2$ analysis, exploring the differential cross-section and the angular coefficients $A_i$ as a function of the transverse momentum of the $Z$ boson $p_{TZ}$. The $\chi^2$ function is defined as follows
\begin{align}
\chi^2=\sum_{ij}\frac{\left(\mathcal{O}_i^{BSM}(p_{TZ,j})-\mathcal{O}_i^{SM}(p_{TZ,j})\right)^2}{(\delta\mathcal{O}_{i}(p_{TZ,j}))^2}\,,
\end{align}
where $\mathcal{O}_i(p_{TZ,j})$ are the observables considered in this analysis for distinct $p_{TZ,j}$ bins. We account for both the binned number of events $N(p_{TZ,j})$  and the angular moments $A_i(p_{TZ,j})$. For the  errors $\delta\mathcal{O}_i(p_{TZ,j})$, we assume $\delta N=\sqrt{N^{SM}+(\epsilon_N N^{SM})^2}$ with systematic uncertainty $\epsilon_N=10\%$~\cite{ATLAS:2021fzm,CMS:2019too}. 
For the angular coefficients, we estimate the statistical uncertainty associated with the measurement of each $A_i(p_{TZ,j})$, performing 100 pseudo-experiments. We consider a random set of $N(p_{TZ,j})$ Monte Carlo events to calculate $A_i(p_{TZ,j})$. We use the standard deviation from the pseudo-experiments to infer the statistical uncertainty on the angular coefficients. The confidence level (C.L.) intervals are defined with 
\begin{align}
1-\text{CL}\ge \int_{\chi^2}^\infty dx p_k(x)\,,
\end{align}
adopting the $\chi^2(c_i/\Lambda^2)$ distribution with $k$ degrees of freedom $p_k(x)$. The CP-conserving effects are evaluated with SM and BSM events samples at NLO QCD. Since the CP-violating operator can only be generated at LO with the UFO model file \texttt{dim6top}, the analysis for this hypothesis accounts for SM and BSM  $t\bar tZ$ and $tZj/\bar{t}Zj$ samples at LO, for consistency.

\begin{table}[t!]
\centering
\begin{tabular}{c|c |c c }
\hline
\hline
\multirow{2}{*}{} &  & $c_i/\Lambda^2$ [TeV$^{-2}$] & $\Lambda/\sqrt{c_i}$ [TeV]\\ 
                 &  & 95\% C.L. bounds & BSM scale \\ 
\hline
\multirow{2}{*}{$c_{tZ}^I$} & linear in $c_i/\Lambda^2$  & [-2.23, 2.23] & 0.67 \\
                          & quadratic in $c_i/\Lambda^2$ & [-1.10, 1.12] & 0.95 \\
\hline
\multirow{2}{*}{$c_{tZ}$} & linear in $c_i/\Lambda^2$    & [-4.63, 4.63] & 0.47 \\
                          & quadratic in $c_i/\Lambda^2$ & [-1.39, 1.26] & 0.89 \\
                          \hline
\multirow{2}{*}{$c_{\phi t}$} & linear in $c_i/\Lambda^2$ & [-4.00, 4.00] & 0.5 \\
                          & quadratic in $c_i/\Lambda^2$ &  [-3.06, 2.94] & 0.58\\
\hline
\multirow{2}{*}{$c_{\phi Q}$} & linear in $c_i/\Lambda^2$  & [-2.61, 2.61] & 0.62 \\
                          & quadratic in $c_i/\Lambda^2$   & [-2.43, 2.83] & 0.64 \\
\hline
\hline
\end{tabular}
\caption{95\% C.L. intervals for the dimension-six operators. The results are presented at linear and quadratic levels in $c_i/\Lambda^2$. The bounds for the CP-conserving operators $(c_{tZ},c_{\phi Q},c_{\phi t})$ are obtained with the observables $(N(p_{tZ}), A_0(p_{TZ}),A_2(p_{TZ}))$. For the operator $c_{tZ}^I$, we also account for the CP-sensitive observable $A_{6}(p_{TZ})$. We assume the HL-LHC at 14~TeV with 3~ab$^{-1}$ of data. 
}
\label{tab:limits}
\end{table}

\begin{figure}[b!]
\centering
\includegraphics[width=0.5\textwidth]{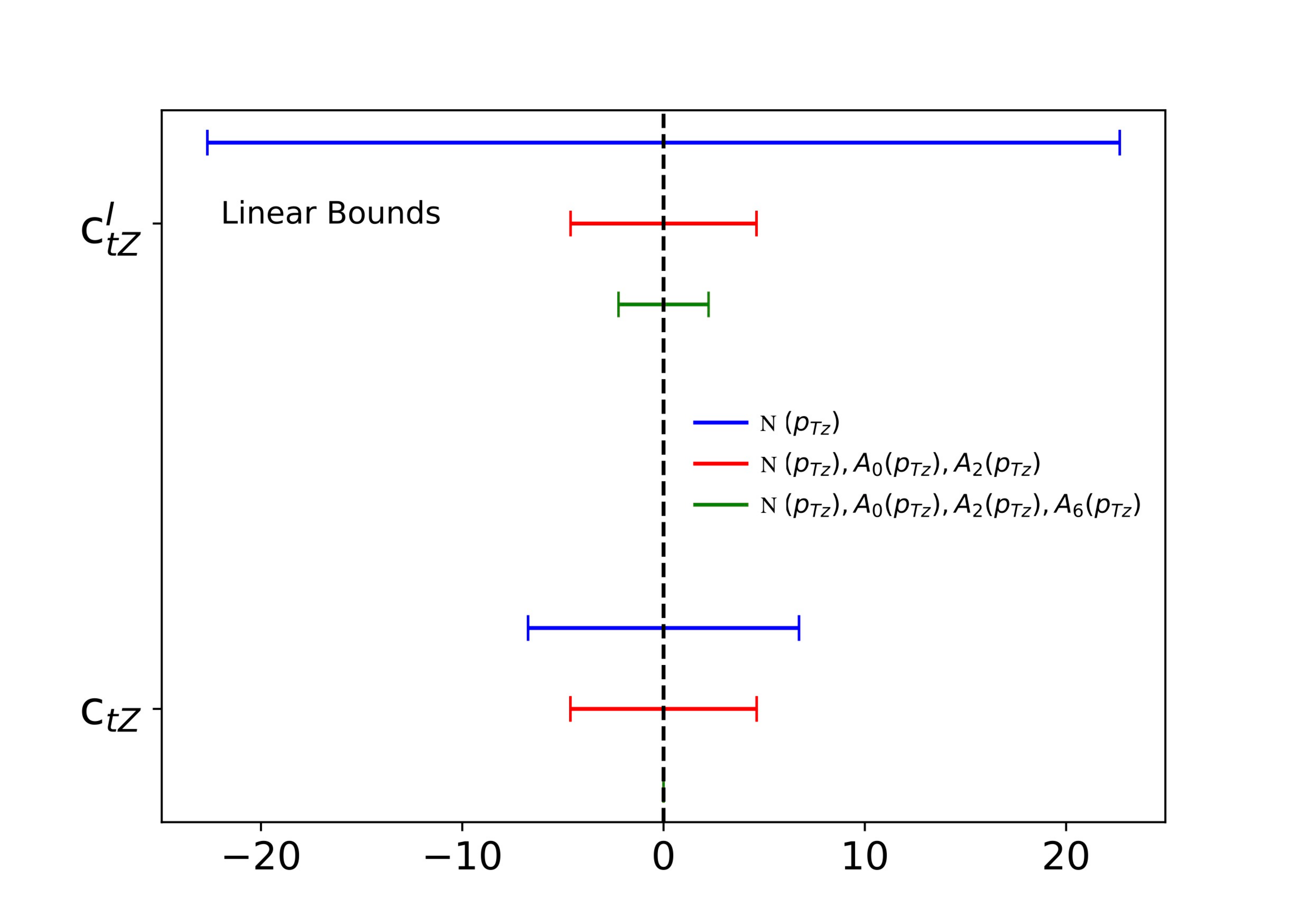}
\caption{95\% C.L. intervals for $c_{tZ}^I$ and $c_{tZ}$ at linear level in $c_i/\Lambda^2$.  The results are shown for three scenarios that differ by the used set of observables: $i)$~$N(p_{p_{TZ}})$ (blue); $ii)$ $N(p_{p_{TZ}}),A_0(p_{TZ}),A_2(p_{TZ})$ (red); and $iii)$ $N(p_{p_{TZ}}),A_0(p_{TZ}),A_2(p_{TZ}),A_6(p_{TZ})$ (green). The latter scenario is only shown for the $c_{tZ}^I$, where $A_6$ displays appreciable sensitivity for the CP-odd effects. See also text and Fig.~\ref{fig:A6}. We assume the HL-LHC at 14~TeV with 3~ab$^{-1}$ of data.
\label{fig:limit}}
\end{figure}

In Table~\ref{tab:limits}, we present the 95\% C.L. constraints on the Wilson coefficients, considering the effects of one BSM operator at a time. We assume the HL-LHC at 14~TeV with 3~ab$^{-1}$ of data.  The results are presented  up to linear  and quadratic level  on the new physics parameters $c_i/\Lambda^2$. To shed light on the extra sensitivity  arising from the angular coefficients, we analyze the $c_{tZ}$ and $c_{tZ}^I$ results in Fig.~\ref{fig:limit} in three scenarios.  The first only explores the binned distribution for the transverse momentum of the $Z$ boson $N({p_{TZj}})$. The second also accounts for the angular coefficients as a function of the energy scale $A_0(p_{TZ,j})$ and $A_2(p_{TZ,j})$. The third one further includes $A_6(p_{TZ,j})$ as an extra probe. We observe that the extra information stored in the angular moments can strongly boost the sensitivity to the Wilson coefficients. Remarkably, while  the analysis of the differential $N(p_{TZ})$ distribution results in no significant sensitivity for $c_{tZ}^I$ at the linear level in the $c_i/\Lambda^2$ expansion, the addition of the angular coefficients $A_i$ result in strong limits at the HL-LHC. In particular, this is due to the new physics effects from the imaginary part of the amplitude that can be probed by the angular coefficient $A_6$.

\section{Conclusion}
\label{sec:conclusion}

In this study, we present a method to augment the new physics sensitivity in searches with the $t\bar{t}Z$ and $tZj/\bar{t}Zj$ processes at the LHC. The proposal explores the accurate measurement of the angular moments for the $Z$ boson, which probes with greater precision the underlying production dynamics. We first access the next to leading order QCD effects for the angular coefficients $A_i$. We observe that the higher order effects can present relevant contributions. Going forward, we parametrize new physics effects in terms of the SM Effective Field theory framework. We observe that the SM and BSM samples display distinct angular coefficients $A_i$. Performing a realistic Monte Carlo study, we show that the angular moments can significantly boost the sensitivity to the Wilson coefficients. In particular, this approach can uncover blind directions to CP-odd operators, leading into sizable sensitivity at the HL-LHC. Remarkably, this proposal only relies on the lepton pair reconstruction, displaying small uncertainties. Hence, it can be promptly incorporated in the ATLAS and CMS analyses.

\begin{acknowledgments}
We would like to thank Rahool Kumar Barman and Zhite Yu for useful discussions. We are also grateful to the authors and maintainers of the TikZ-Feynman package~\cite{Ellis:2016jkw} that was used to generate the Feynman diagrams in this work. RMA and DG thank the U.S.~Department of Energy for the financial support, under grant number DE-SC 0016013. Some computing for this project was performed at the High Performance Computing Center at Oklahoma State University, supported in part through the National Science Foundation grant OAC-1531128.
\end{acknowledgments}

\bibliography{references}
\end{document}